\documentclass[aps,showpacs,two column,showkeys,superscriptaddress,preprintnumbers]{revtex4-2}
\usepackage{epsfig,amssymb,subfigure,bm,dsfont}
\usepackage{chngcntr}
\usepackage{amsmath}
\usepackage{gensymb}
\usepackage{txfonts}
\usepackage[titletoc]{appendix}
\usepackage{amssymb}
\usepackage{array}
\usepackage{mathrsfs}
\usepackage{subfigure,overpic}
\usepackage{dcolumn}
\usepackage{graphicx}
\usepackage{epstopdf}
\usepackage{float}
\usepackage{balance}
\usepackage{lipsum}
\usepackage[colorlinks,
            linkcolor=blue,
            anchorcolor=blue,
            citecolor=blue,
            urlcolor=blue]{hyperref}
\usepackage{setspace}
\usepackage{multirow}
\usepackage{makecell}
\usepackage[normalem]{ulem} % 波浪线包
\usepackage{xcolor}
\usepackage{pifont}
\usepackage{graphicx}
\usepackage{tikz}     % 更高级的控制
\usepackage[capitalize]{cleveref}
\usetikzlibrary{decorations.pathmorphing}
\definecolor{myblue}{rgb}{0,0,1}
\definecolor{myred}{rgb}{1,0,0}
\definecolor{myblack}{rgb}{0,0,0}
\definecolor{mymagenta}{rgb}{1,0,1}

\begin{document}

\title{RKKY signatures as a probe for intrinsic magnetism and AI/QAH phase discrimination in MnBi$_2$Te$_4$ films}

\author{Ya-Xi Li}
\affiliation{Guangdong Basic Research Center of Excellence for Structure and Fundamental Interactions of Matter, Guangdong Provincial Key Laboratory of Quantum Engineering and Quantum Materials, School of Physics, South China Normal University, Guangzhou 510006, China}
\affiliation{Guangdong-Hong Kong Joint Laboratory of Quantum Matter, Frontier Research Institute for Physics, South China Normal University, Guangzhou 510006, China}
\author{Zi-Jian Chen}
\affiliation{Guangdong Basic Research Center of Excellence for Structure and Fundamental Interactions of Matter, Guangdong Provincial Key Laboratory of Quantum Engineering and Quantum Materials, School of Physics, South China Normal University, Guangzhou 510006, China}
\affiliation{Guangdong-Hong Kong Joint Laboratory of Quantum Matter, Frontier Research Institute for Physics, South China Normal University, Guangzhou 510006, China}
\author{Rui-Qiang Wang}
\affiliation{Guangdong Basic Research Center of Excellence for Structure and Fundamental Interactions of Matter, Guangdong Provincial Key Laboratory of Quantum Engineering and Quantum Materials, School of Physics, South China Normal University, Guangzhou 510006, China}
\affiliation{Guangdong-Hong Kong Joint Laboratory of Quantum Matter, Frontier Research Institute for Physics, South China Normal University, Guangzhou 510006, China}
\author{Ming-Xun Deng}
\affiliation{Guangdong Basic Research Center of Excellence for Structure and Fundamental Interactions of Matter, Guangdong Provincial Key Laboratory of Quantum Engineering and Quantum Materials, School of Physics, South China Normal University, Guangzhou 510006, China}
\affiliation{Guangdong-Hong Kong Joint Laboratory of Quantum Matter, Frontier Research Institute for Physics, South China Normal University, Guangzhou 510006, China}
\author{Mou Yang}
\affiliation{Guangdong Basic Research Center of Excellence for Structure and Fundamental Interactions of Matter, Guangdong Provincial Key Laboratory of Quantum Engineering and Quantum Materials, School of Physics, South China Normal University, Guangzhou 510006, China}
\affiliation{Guangdong-Hong Kong Joint Laboratory of Quantum Matter, Frontier Research Institute for Physics, South China Normal University, Guangzhou 510006, China}
\author{Hou-Jian Duan}
\email{dhjphd@163.com}
\affiliation{Guangdong Basic Research Center of Excellence for Structure and Fundamental Interactions of Matter, Guangdong Provincial Key Laboratory of Quantum Engineering and Quantum Materials, School of Physics, South China Normal University, Guangzhou 510006, China}
\affiliation{Guangdong-Hong Kong Joint Laboratory of Quantum Matter, Frontier Research Institute for Physics, South China Normal University, Guangzhou 510006, China}

\begin{abstract}
We present a systematic study of the Ruderman-Kittel-Kasuya-Yosida (RKKY) interaction in MnBi$_2$Te$_4$ films under both dark and illuminated conditions. In the dark, the intrinsic magnetism of MnBi$_2$Te$_4$ is shown to yield a stronger anisotropic RKKY spin model compared to nonmagnetic topological insulators, providing a clear signature for differentiating these systems. Furthermore, key band properties---such as energy gap, band degeneracy/splitting, and topological deformations of the Fermi surface---imprint distinct signatures on the RKKY interaction, enabling clear discrimination between axion insulators (AI) and quantum anomalous Hall (QAH) insulators in even- and odd-septuple-layer (SL) films. This discrimination manifests in multiple ways: through the Fermi-energy dependence or spatial oscillations of the interaction for impurities on the same surface, or via the presence versus absence of spin-frustrated terms for those on different surfaces. Under off-resonant circularly polarized light, additional phase-transition-related fingerprints also emerge to distinguish these two phases, such as sign reversals of spin-frustrated terms in even-SL films versus chirality-selective double-dip structures of collinear RKKY components in odd-SL films. Overall, this work establishes RKKY interactions as a sensitive magnetic probe for distinguishing between AI phase (even-SL) and QAH phase (odd-SL), thereby complementing conventional electrical measurements while providing new insights into the influence of intrinsic magnetism on the surface-state band structure.
\end{abstract}

\maketitle

%%\altaffiliation{Electronic address: yang.mou@hotmail.com}

%%%%%%%%%%%%%%%%%%%%%%%%%%%%%%%%%%%%%

\section{Introduction}
The interplay between nonmagnetic topological materials and magnetism enables the realization of novel topological phases, such as quantum anomalous Hall (QAH) insulators, axion insulators (AI), and Weyl semimetals \cite{1,2,3,4,5,6,7,8,9,10,11,13}. These systems exhibit remarkable transport phenomena, including the dissipationless chiral edge states in QAH insulators \cite{6,7,8}, the quantized magnetoelectric effects in topological axion states \cite{9,10,11}, and the non-Abelian statistics of Majorana fermions \cite{13}. Such unique properties render magnetic topological states highly promising for applications in spintronics and topological quantum computing. Conventionally, these states are achieved by incorporating magnetic dopants into host materials. A landmark in this approach was the prediction and subsequent observation of the QAH effect in Cr-doped (Bi,Sb)$_2$Te$_3$ thin films \cite{7,8,14}. However, this method relies on extrinsic magnetism, making the resultant topological properties highly sensitive to the precise chemical composition and typically limited to very low temperatures \cite{15}. These inherent drawbacks pose a formidable challenge for engineering diverse phases through doping, thus intensifying the search for a new class of intrinsic magnetic topological materials, where magnetism is an inherent property of the crystal lattice itself.
\par
Among such intrinsic candidates, MnBi$_2$Te$_4$ has emerged as a canonical platform, attracting significant interest \cite{16,cpl1}. Its layered crystal structure, composed of septuple layers (SLs) with van der Waals bonding and an A-type antiferromagnetic order, provides the foundation for unique quantum phenomena when confined to films. In the few-SL limit, a fundamental dichotomy arises: both the magnetic order and the resulting emergent topological states are governed entirely by the parity of the SL count \cite{17,18,19,20}. Specifically, odd-SL films exhibit ferromagnetic (FM) order, characterized by parallel magnetization on their top and bottom surfaces. This FM configuration gives rise to the QAH state, identified by a Chern number $C = +1$ and a quantized Hall conductance $e^{2}/h$. In contrast, even-SL films maintain antiferromagnetic (AFM) order with antiparallel surface magnetizations, stabilizing the so-called AI state. The distinctive feature of this AI---setting it apart from a trivial insulator---is its quantized magnetoelectric coefficient $\theta =\pi$ \cite{10,21,22,23}, even as it exhibits a zero Hall plateau ($C = 0$). Owing to this striking thickness dependence and intrinsic magnetism, MnBi$_{2}$Te$_{4}$ films stand as a highly promising platform for fundamental research and future device applications \cite{20}.
\par
A key challenge in studying MnBi$_2$Te$_4$ films lies in the unambiguous identification of their distinct topological phases. Specifically, initial breakthroughs demonstrated the material's potential: the quantized QAH state was realized in an odd-SL film \cite{23_1}, and the AI state with a zero-Hall plateau was reported in an even-SL film \cite{20}, both at zero magnetic field. However, subsequent studies have revealed that electrical transport measurements alone are insufficient to reliably distinguish these states \cite{24}. For instance, a nominal 5-SL (odd) device was found to exhibit vanishing Hall resistance and high longitudinal resistance---signatures previously associated with the even-SL AI state \cite{20}---highlighting the ambiguity. This practical difficulty arises because the dissipationless edge current expected in an ideal QAH phase can be disrupted or shunted by various imperfections, allowing transport to become dominated by the insulating bulk or other dissipative channels \cite{26}. Consequently, the characteristic transport signatures of a topological phase can be masked, making the sole reliance on resistivity and Hall measurements problematic for definitive phase identification. This ambiguity underscores the critical need for complementary, non-transport probes that can access the topological nature of the surface states. 
\par
Interestingly, such complementary probes can be sought in two distinct regimes: in unperturbed (dark) systems and in systems under a dynamical perturbation such as circularly polarized light (CPL). For the former, the key lies in selecting a probing mechanism that does not perturb the system itself. For the latter, the situation is more intricate because CPL induces a rich variety of topological phase transitions in MnBi$_2$Te$_4$ films \cite{45,46}. Fortunately, these transitions depend critically on the film type (even- vs. odd-SL). This dependence suggests that tracing these topological phase transitions may offer a potential route to differentiate between AI phase (even-SL) and QAH phase (odd-SL). Together, this dual-pathway (dark and illuminated) strategy can construct a more comprehensive probing scheme.
\par

To implement this dual-pathway strategy, we investigate the RKKY interaction as a unified magnetic probe in MnBi$_2$Te$_4$ films. This proposal rests on two pillars. First, for magnetic impurities placed on the film surface with a relatively large separation, the induced RKKY interaction is weak and leaves the host's band structure unperturbed \cite{44_3}, making it suitable for probing the unperturbed (dark) system. Second, the RKKY interaction has been demonstrated to be a powerful tool for characterizing band structures and topological properties (even in perturbed systems) \cite{27_1,27,28_1,28,29,30,31,32,33,34,35,36,37,38,39,40,41,42,43,44,44_1,44_2}. By analyzing this interaction within the established low-energy model \cite{45}, we pursue two primary objectives: (i) to extract characteristic RKKY signatures in the dark that distinguish between the nonmagnetic topological insulator, AI (even-SL), and QAH (odd-SL) phases, thereby addressing the experimental ambiguities; and (ii) to track its evolution through the photoinduced topological phase transitions under CPL, thereby identifying additional magnetic fingerprints to distinguish between the AI and QAH phases. This dual-signature approach establishes the RKKY interaction as a comprehensive magnetic probe that complements conventional electrical measurements.

\par
The paper is structured as follows. Sec. \ref{model} presents the low-energy model for the surface states of MnBi${_2}$Te${_4}$ films, the rich topological phase transitions induced by CPL, and the method for calculating the RKKY interaction. Sec. \ref{results} investigates three key aspects: (a) the influence of intrinsic magnetism on the RKKY spin model and the identification of signatures to distinguish MnBi${_2}$Te${_4}$ from nonmagnetic topological insulators; (b) the evolution of the collinear RKKY components with Fermi energy and its spatial oscillatory behavior, together with the characteristics of spin-frustrated terms, which serve to differentiate between AI phase (even-SL) and QAH phase (odd-SL); and (c) the variation of the RKKY amplitude with light parameters, enabling the extraction of additional phase-transition-related signatures for distinguishing AI phase (even-SL) and QAH phase (odd-SL). A summary is provided in Sec. \ref{summary}.

\section{Models and method}
\label{model}
\subsection{The effective model}
\begin{figure}[tbh]
\centering \includegraphics[width=0.44\textwidth]{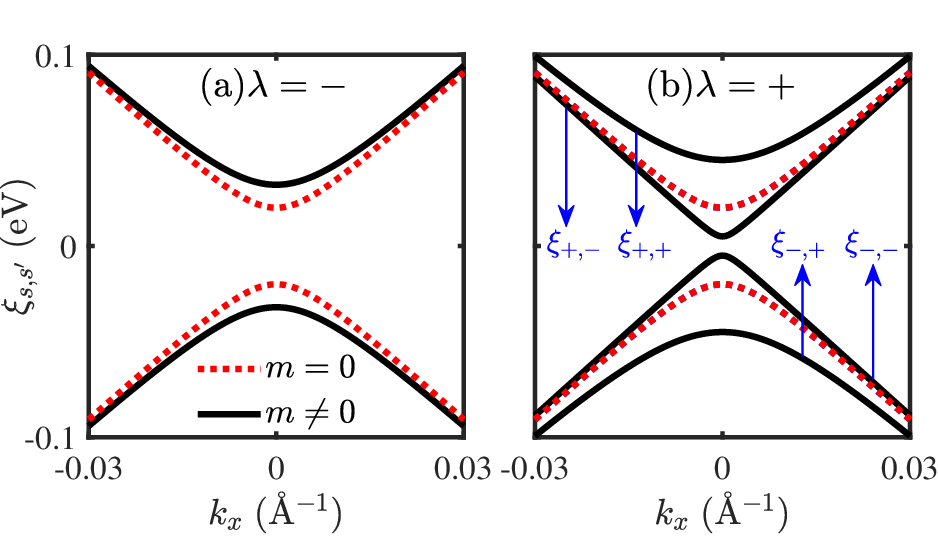}
\caption{$k_x$-axis dispersion for (a) even-SL ($\lambda = -$) and (b) odd-SL ($\lambda = +$) MnBi$_2$Te$_4$ films at different Zeeman coupling strengths ($m = 0$, $0.025$ eV). Other parameters are $\Delta = 0.02$ eV and $v = 2.95\ \mathrm{eV \cdot \r{A}}$.}
\label{fig1}
\end{figure}
The effective Hamiltonian describing the surface states of MnBi$_2$Te$_4$ films---a model captured in Refs. \cite{16,17,18,19,20} and formalized in Ref. \cite{45}---is given by
\begin{equation}\label{eq1}
H_{0}(\mathbf{k})=\left(
\begin{array}{cc}
h_{+,+}(\mathbf{k}) & \Delta \sigma _{0} \\
\Delta \sigma _{0} & h_{-,\lambda }(\mathbf{k})%
\end{array}%
\right) .
\end{equation}%
This Hamiltonian effectively represents two coupled Dirac cones originating from the top and bottom surfaces of the film. In this expression, the Hamiltonian for an individual Dirac cone is given by $h_{s_i, \lambda}(\mathbf{k}) = s_i v (\mathbf{k} \times \boldsymbol{\sigma})_z + \lambda m \sigma_z$, where $s_i = \pm 1$ specifies the helicity of the Dirac cone, and $\lambda = \pm$ distinguishes between odd- ($\lambda = +$) and even-SL ($\lambda = -$) films. Here, $\boldsymbol{\sigma} = (\sigma_x, \sigma_y, \sigma_z)$ denotes the vector of Pauli matrices in spin space, with $\sigma_0$ representing the identity matrix.  In this model, the parameters are defined as follows: $v$ corresponds to the Fermi velocity, $\Delta$ quantifies the coupling between the two surface Dirac cones induced by finite-size effects, and $m$ represents the strength of intrinsic magnetism, which arises from the Zeeman coupling associated with time-reversal-symmetry-breaking magnetic moments. When $m \neq 0$, the sign of $\lambda$---which is determined by the parity of the SL count---governs the selection between the AI phase ($\lambda = -$) and the QAH phase ($\lambda = +$). It is worth noting that the model reduces to that of a nonmagnetic topological insulator film for $m=0$. In addition, the validity of the surface-state model in Eq.~(\ref{eq1}) critically depends on the film thickness: the film must be neither too thin nor too thick. Specifically, to maintain the QAH state in odd-SL films, Ref.~\cite{19} explicitly points out that the thickness cannot be 1~SL. The currently available theoretical studies \cite{19,46} and experimentally fabricated MnBi$_2$Te$_4$ films \cite{17} cover a thickness range of 2--9~SL. Within this thickness range, the model in Eq.~(\ref{eq1}) provides a faithful description of the QAH and AI states reported for even- and odd-SL films in Refs.~\cite{17,19,46}. Furthermore, our parameter choices and the bandwidth ($\approx 0.2$~eV) of the low-energy model are consistent with those in Ref.~\cite{45}. The Fermi energies (0.06~eV without light and 0~eV with light) used in our RKKY calculations are well below this bandwidth, which ensures the reliability of our calculations and the associated analysis.

\begin{figure}[tbh]
\centering \includegraphics[width=0.44\textwidth]{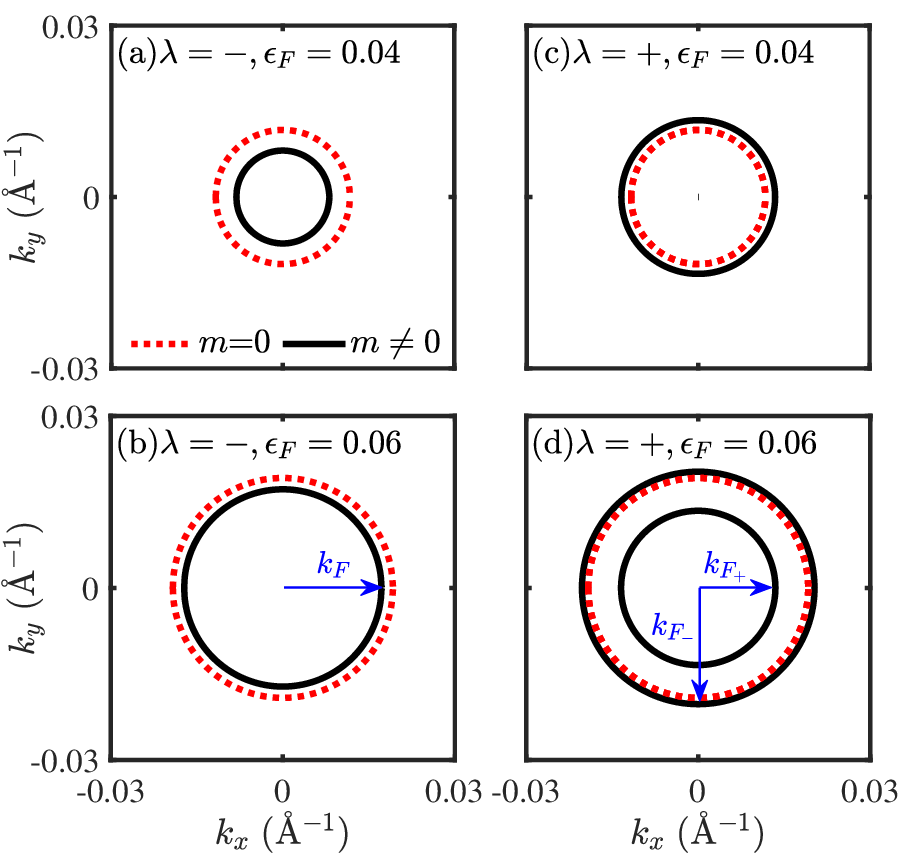}
\caption{Fermi surfaces of (a,b) even-SL ($\lambda=-$) and (c,d) odd-SL ($\lambda=+$) MnBi$_2$Te$_4$ thin films for different Zeeman coupling strengths $m = 0, 0.025$ eV. The Fermi energy is set to $\epsilon_F = 0.04$ eV in (a,c) and $0.06$ eV in (b,d). Other parameters are identical to those in Fig. \ref{fig1}. Labels $k_F$ and $k_{F_\pm}$ in (b,d) denote the Fermi wave numbers for the even- and odd-SL films, respectively.}
\label{fig2}
\end{figure}

By diagonalizing the Hamiltonian in Eq. (\ref{eq1}), we obtain the energy dispersion
\begin{equation}\label{eq2}
\xi_{s,s'}(m, \lambda)=s\sqrt{m^{2}+k^{2}v^{2}+\Delta ^{2}+s^{\prime }\sqrt{%
2\Delta ^{2}m^{2}(1+\lambda )}},
\end{equation}%
where $s = \pm$ labels the conduction-band and valence-band doublets, respectively, and $s^{\prime} = \pm$ indexes the two subbands within either doublet. At $m=0$, time-reversal ($\mathcal{T}$) symmetry ensures the degeneracy between the two subbands $\xi_{s,+}$ and $\xi_{s,-}$. The introduction of intrinsic magnetism ($m \neq 0$) in even-SL ($\lambda=-$) films breaks $\mathcal{T}$ symmetry; however, the band degeneracy is still preserved by the combined symmetry of inversion ($\mathcal{P}$) and $\mathcal{T}$ \cite{26,46}. In contrast to the $m=0$ case, the bands experience a momentum-dependent shift: the conduction band moves upward while the valence band downward. This shift is most pronounced at $k_{x,y}=0$, which enlarges the band gap from $\xi_g(m=0)=2\Delta$ to $\xi_g(m \neq 0,\lambda=-)=2\sqrt{m^2+\Delta^2}$, as illustrated in Fig. \ref{fig1}(a). Conversely, introducing magnetism ($m \neq 0$) in odd-SL ($\lambda=+$) films breaks $\mathcal{PT}$ symmetry, which lifts the band degeneracy, resulting in split subbands where $\xi_{s,+} \neq \xi_{s,-}$, as depicted in Fig. \ref{fig1}(b). Examining the conduction band in detail, one subband, $\xi_{+,+}(m\neq0,\lambda=+)$, rises above the original $\xi_{+,s'}(m=0)$ level (represented by the dashed line in Fig. \ref{fig1}(b)), while the other, $\xi_{+,-}(m\neq0,\lambda=+)$, falls below it. Consequently, the band gap narrows to $\xi_g(m \neq 0,\lambda=+)=2|m-\Delta|$.
\par
In short, the introduction of $m$ significantly modifies the band structure of both even- and odd-SL films, demonstrating a clear dependence on the SL count. This dependence is further reflected in distinct deformations of the Fermi surface. As shown in Figs. \ref{fig2}(a) and \ref{fig2}(b), the Fermi surface in even-SL ($\lambda=-$) films remains a single circle for different Fermi energies. This circle is an enlarged version of the $m=0$ case, and the degree of enlargement decreases as the Fermi energy increases. Thus, varying the Fermi energy does not alter the topology of the Fermi surface. In contrast, for odd-SL ($\lambda=+$) films, varying the Fermi energy drives a topological deformation of the Fermi surface, i.e., a Lifshitz transition [Figs. \ref{fig2}(c) and \ref{fig2}(d)]. Specifically, at low Fermi energies, the Fermi surface is a single circle [Fig. \ref{fig2}(c)], while at higher energies it consists of two concentric circles [Fig. \ref{fig2}(d)]---one contracted and the other expanded relative to the original $m=0$ circular Fermi surface. This transition originates entirely from the unique band splitting shown in Fig. \ref{fig1}(b) and represents a key band characteristic that distinguishes QAH insulators from both AI and nonmagnetic topological insulators.
\par
\begin{figure}[tbh]
\centering \includegraphics[width=0.44\textwidth]{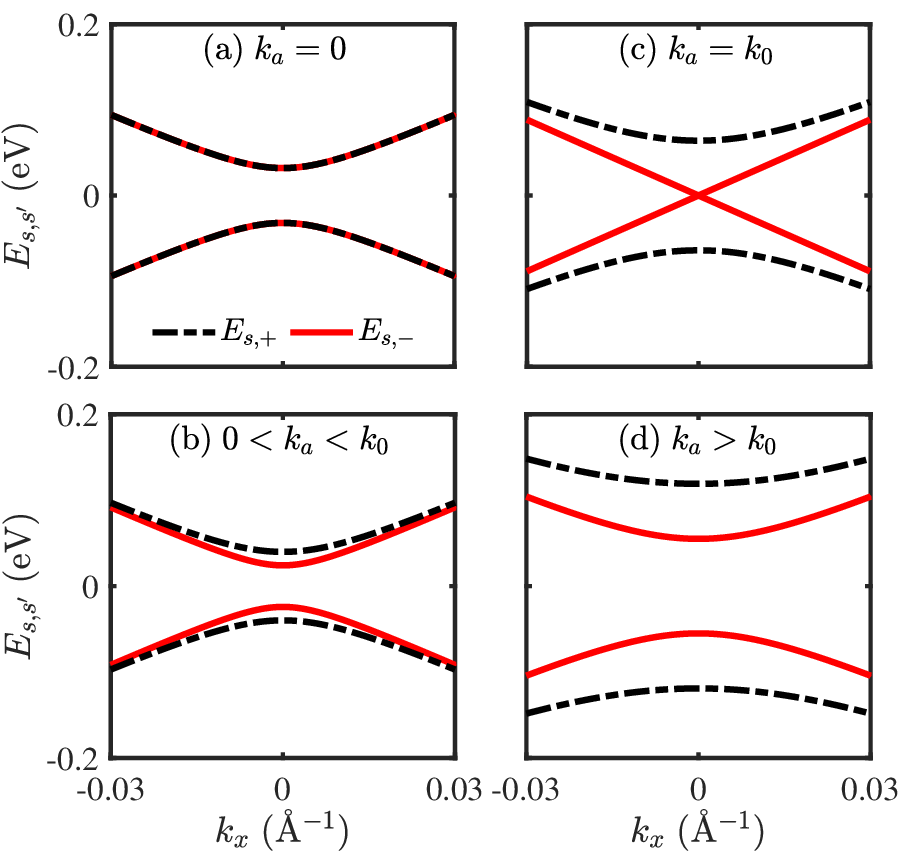}
\caption{Evolution of the $k_x$-axis dispersion for even-SL ($\lambda=-$) MnBi$_2$Te$_4$ films at different light parameter $k_a$: (a) $k_a=0$, (b) $k_a<k_0$ (where $k_{0}=\sqrt[4]{\hbar^2 \omega^2\left(m^{2}+\Delta ^{2}\right)}/v$; here $k_a=0.03~{\rm \AA^{-1}}$), (c) $k_a=k_0$, and (d) $k_a>k_0$ ($k_a=0.1~{\rm \AA^{-1}}$). Results are shown for right-handed ($\eta=+$) CPL; identical dispersion is obtained for left-handed ($\eta=-$) polarization.}
\label{fig3}
\end{figure}

\begin{figure*}[tbh]
\centering \includegraphics[width=0.99\textwidth]{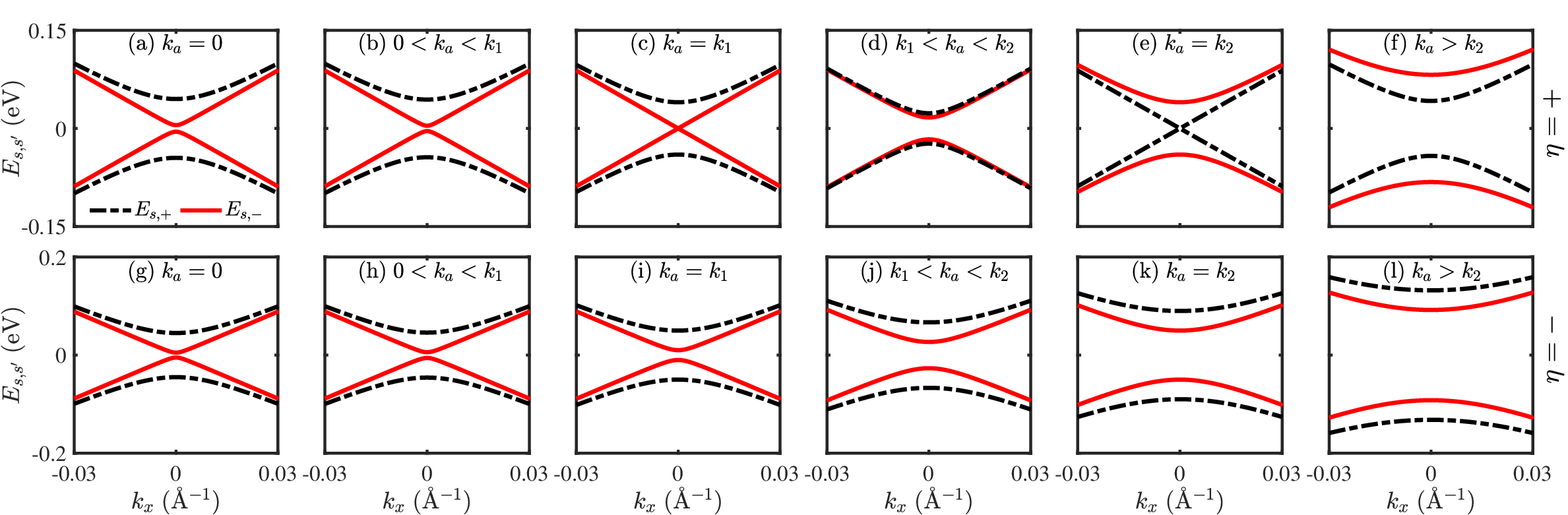}
\caption{$k_x$-axis dispersion for odd-SL ($\lambda=+$) MnBi$_2$Te$_4$ films at different light parameter $k_a$ (where $k_{1}=\sqrt{\hbar \omega(m-\Delta)}/v$ and $k_2 = \sqrt{\hbar \omega (m + \Delta)}/v$): (a,g) $k_a=0$, (b,h) $k_a<k_1$ ($k_a=0.01~{\rm \AA^{-1}}$), (c,i) $k_a=k_1$, (d,j) $k_1<k_a<k_2$ ($k_a=0.05~{\rm \AA^{-1}}$), (e,k) $k_a=k_2$, and (f,l) $k_a>k_2$ ($k_a=0.1~{\rm \AA^{-1}}$). The top row (a-f) and bottom row (g-l) correspond to right- ($\eta=+$) and left-handed ($\eta=-$) CPL, respectively.}
\label{fig4}
\end{figure*}
Under irradiation by CPL, MnBi$_2$Te$_4$ films exhibit a rich variety of topological phase transitions, as documented in Refs. \cite{45,46}. In contrast to conventional photoinduced topological transitions, these phases depend critically on both the chirality of the circular polarization and the number of SL. To systematically track how the energy bands and topological indices evolve with the parameters of the light, we compute the effective Hamiltonian under illumination. We begin by assuming normal incidence of CPL on the film surface. The time-dependent Hamiltonian is obtained via Peierls substitution: $\mathbf{k} \rightarrow \mathbf{k} + e\mathbf{A}/\hbar$, with the vector potential $\mathbf{A}(t) = A_0[0, \cos(\omega t), \eta\sin(\omega t)]$ and period $T = 2\pi/\omega$. Here, $\eta = +$ (or $-$) denotes right-handed (or left-handed) polarization, and $A_0$ is the amplitude of the vector potential. Applying Floquet theory \cite{47_1,47_2,47,48,49} under the off-resonant condition $\hbar\omega \gg BW$ (with $\hbar\omega = 1\ \mathrm{eV}$ and the bandwidth $BW = 0.2\ \mathrm{eV}$), the photoinduced correction to the Hamiltonian takes the form
\begin{equation}\label{eq3}
H(\mathbf{k})=V_{0}+\sum_{n\geq 1}\frac{\left[ V_{+n},V_{-n}\right] }{n\hbar
\omega}+O\left(\frac{1}{\omega ^{2}}\right),
\end{equation}
where $V_{n}=\frac{1}{T}%
\int_{0}^{T}H_0(\mathbf{k}+e\mathbf{A}/\hbar)e^{-in\hbar \omega t}dt$. After some algebraic calculations, $V_n$ can be solved as
 \begin{eqnarray}\label{eq4}
 \begin{split}
 V_0&=H_{0}(\mathbf{k}),\\
 V_{+1}&=  \frac{vk_a}{2}\left(
 \begin{array}{cc}
 	\sigma_{y}-i\eta \sigma_x  & 0 \\
 	0 &  -\sigma_{y}+i\eta \sigma_x
 \end{array}%
 \right) ,
 \end{split}
 \end{eqnarray}
where $k_a=eA_{0}/\hbar$. Other Floquet sidebands follow as $V_{-1}=V_{+1}^{\dagger}$ and $V_n=0$ for $n\geq2$. Substituting these results into Eq. (\ref{eq3}) yields the effective Hamiltonian
\begin{equation}\label{eq5}
H(\mathbf{k})=H_{0}(\mathbf{k})-\eta m_{\omega }\sigma _{z},
\end{equation}
where $m_{\omega }=v^{2}k_{a}^{2}/(\hslash \omega )$.
\par
The diagonalization of $H(\mathbf{k})$ in Eq. (\ref{eq5}) leads to the photon-dressed energy dispersion:
 \begin{eqnarray}\label{eq6}
 \begin{split}
 E_{s,s'}(\lambda=-)&=s\sqrt{k^{2}v^{2}+\left(m_{\omega}+s^{\prime }\sqrt{\Delta ^{2}+m^{2}}\right)^2},\\
E_{s,s'}(\lambda=+)&=s\sqrt{k^{2}v^{2}+\left[\Delta +s^{\prime }\left(m-\eta m_{\omega }\right)\right]^{2}}.
 \end{split}
 \end{eqnarray}
Based on Eqs. (\ref{eq6}), we plot the energy bands for various values of $k_a$ in Figs. \ref{fig3} and \ref{fig4} to track the evolution of the dispersion. For even-SL ($\lambda=-$) films, as $k_a$ increases from zero, the surface-state bands undergo a gap closing and reopening at $k_a = k_0$ ($k_{0}=\sqrt[4]{\hbar^2 \omega^2\left(m^{2}+\Delta ^{2}\right)}/v$), as shown in Fig. \ref{fig3}. This evolution is independent of the chirality of the circular polarization, resulting in identical bands for both $\eta=+$ and $\eta=-$. In contrast, the response of odd-SL ($\lambda=+$) films depends strongly on the polarization chirality. Specifically, for right-handed polarization ($\eta=+$), increasing $k_a$ induces successive gap closings and reopenings at $k_a = k_1$ and $k_a = k_2$, with $k_{1}=\sqrt{\hbar \omega(m-\Delta)}/v$ and $k_2 = \sqrt{\hbar \omega (m + \Delta)}/v$. The first gap closing at $k_1$ is dominated by the $E_{s,-}$ band (red solid lines in Fig. \ref{fig4}(c)), whereas the second at $k_2$ is dominated by the $E_{s,+}$ band (black dashed lines in Fig. \ref{fig4}(e)). For left-handed polarization ($\eta=-$), however, increasing $k_a$ monotonically enlarges the gap without any closing or reopening. Since a gap closing and reopening can potentially induce a topological phase transition, we thus expect rich topological transitions in even-SL ($\lambda=-$) films, as well as in odd-SL ($\lambda=+$) films under right-handed polarization.
\begin{figure}[tbh]
\centering \includegraphics[width=0.48\textwidth]{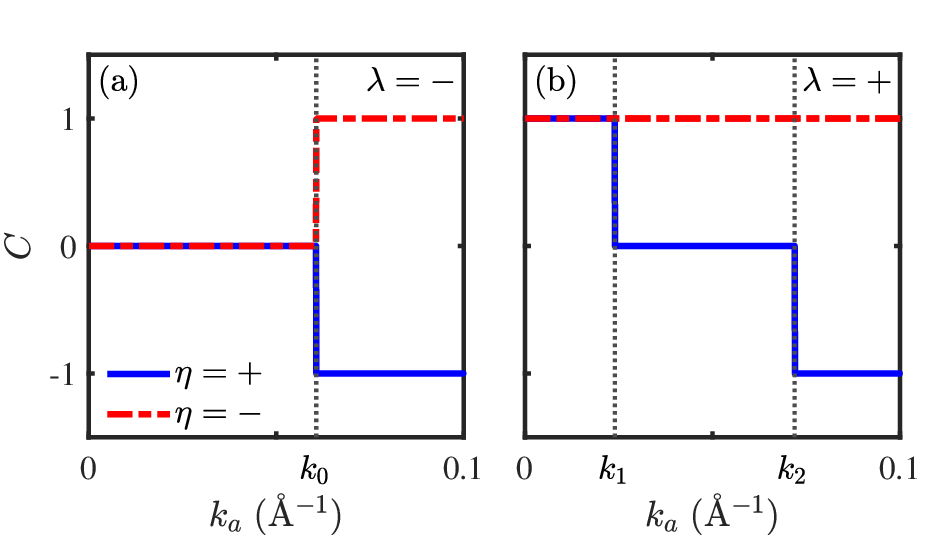}
\caption{Chern number vs. light parameter $k_a$ for (a) even-SL ($\lambda = -$) and (b) odd-SL ($\lambda = +$) MnBi$_2$Te$_4$ films under both right- ($\eta=+$) and left-handed ($\eta=-$) CPL. Parameters: $m=0.025$ eV; others as in Fig. \ref{fig1}.}
\label{fig5}
\end{figure}
\par
To further clarify and confirm the topological phase transitions in MnBi$_2$Te$_4$ films under CPL, we evaluate the Chern number $C$, which is defined as
\begin{equation}\label{eq7}
C=\frac{1}{2\pi }\underset{n}{\sum }\int_{BZ}dk\Omega _{xy}^{n},
\end{equation}
where the Berry curvature $\Omega _{xy}^{n}$ for the $n$th band is expressed as \cite{50}
\begin{equation}\label{eq8}
\Omega _{xy}^{n}(\mathbf{k})=i\underset{m\neq n}{\sum }\frac{\left\langle
u_{n}|\frac{\partial H}{\partial k_{x}}|u_{m}\right\rangle \left\langle
u_{m}|\frac{\partial H}{\partial k_{y}}|u_{n}\right\rangle
-(x\leftrightarrow y)}{(E_{n}-E_{m})^{2}}.
\end{equation}
Here, $m$ is a band index, while $E_n$ and $|u_n\rangle$ denote the eigenvalue and eigenstate of the $n$-th band, respectively. We set the Fermi energy to zero so that the Chern number $C$ sums over all occupied valence bands, which correspond to the two valence subbands shown in Figs. \ref{fig3} and \ref{fig4}. Substituting Eq. (\ref{eq8}) into Eq. (\ref{eq7}) and performing the integration, we arrive at the following explicit results:
\begin{eqnarray}\label{eq9}
\begin{split}
C(\lambda =-)&=\frac{\mathrm{sgn}(\sqrt{\Delta ^{2}+m^{2}}+\eta m_{\omega })-\mathrm{sgn}(\sqrt{\Delta ^{2}+m^{2}}-\eta m_{\omega })}{2},\\
 C(\lambda =+)&=\frac{\mathrm{sgn}(m+\eta m_{\omega }+\Delta )+\mathrm{sgn}(m+\eta m_{\omega }-\Delta )}{2}.
\end{split}
\end{eqnarray}
\par
Using Eq. (\ref{eq9}), we trace the evolution of the Chern number with the light parameter $k_a$, as shown in Fig. \ref{fig5}. In the even-SL ($\lambda=-$) case [Fig. \ref{fig5}(a)], once $k_a$ exceeds the critical value $k_0$, the system transitions from the original AI phase ($C=0$) into a QAH phase. The Chern number of this resulting QAH phase depends on the polarization chirality: it takes $C=-1$ for $\eta=+$ and $C=1$ for $\eta=-$. For the odd-SL ($\lambda=+$) films under right-handed polarization ($\eta=+$), the system undergoes two phase transitions as $k_a$ increases. As illustrated in Fig. \ref{fig5}(b), it evolves successively from a QAH phase ($C=+1$) to a normal insulator (NI) phase ($C=0$), and finally to another QAH phase ($C=-1$). The transition points occur precisely at $k_a = k_1$ and $k_a = k_2$, which correspond to the gap-closing cases shown in Figs. \ref{fig4}(c) and \ref{fig4}(e), respectively. In contrast, under left-handed polarization ($\eta=-$), the system remains in the QAH phase for all $k_a$, as no gap closure and reopening occurs (see Fig. \ref{fig4}). These findings are fully consistent with the results reported in Ref. \cite{45}. Building on this optically controllable platform, we next aim to extract the magnetic signatures of these chirality-dependent, photon-induced topological phase transitions.

\subsection{The RKKY interaction}
We model the RKKY interaction in a MnBi$_2$Te$_4$ film by considering two magnetic impurities placed on its surfaces, located at positions $\mathbf{r}_i$ and $\mathbf{r}_j$, respectively. The total Hamiltonian of the system, which includes the spin-exchange interaction between the impurities and the host electrons within the $s$-$d$ model, is given by
\begin{eqnarray}\label{eq10}
\begin{split}
H^\prime&= H + H_{\text{int}} ,\\
&= H - J_c \mathbf{S}_i \cdot \mathbf{s}_i - J_c \mathbf{S}_j \cdot \mathbf{s}_j,
\end{split}
\end{eqnarray}
where $J_c$ denotes the strength of the exchange coupling, $\mathbf{S}_i$ represents the spin of the impurity at site $\mathbf{r}_i$, and $\mathbf{s}_i = \frac{1}{2} c^\dagger_{i\alpha} \boldsymbol{\sigma}_{\alpha\beta} c_{i\beta}$ is the spin of the host electrons at the same site. Since the two impurities couple indirectly via the itinerant electrons, the resulting effective exchange interaction between them is the RKKY interaction. In the weak-coupling limit where $J_c$ is sufficiently small, $H_{\text{int}}$ can be treated as a perturbation. Applying standard second-order perturbation theory \cite{51,52,53,54} in $J_c$, the explicit form of this RKKY interaction is derived as:
\begin{equation}\label{eq11}
H_{R}^{\alpha\beta}=-\frac{J_{c}^{2}}{\pi }\mathrm{Im}\int_{-\infty }^{\epsilon
_{F}}d\epsilon \mathrm{Tr}\left[ (\mathbf{S}_{1}\cdot \sigma )G_{\alpha\beta}(\mathbf{%
R},\epsilon)(\mathbf{S}_{2}\cdot \sigma )G_{\beta\alpha}(-\mathbf{R},\epsilon
)\right] ,
\end{equation}%
where $\mathbf{R} = \mathbf{r}_i - \mathbf{r}_j$, $\epsilon_F$ is the Fermi energy, and $G_{\alpha\beta}(\pm\mathbf{R}, \epsilon)$ denotes a matrix element of the retarded Green's function $G(\pm\mathbf{R}, \epsilon)$ associated with the unperturbed Hamiltonian $H$ in real space. The subscripts $\alpha,\beta \in {t, b}$ indicate whether an impurity is located on the top ($t$) or bottom ($b$) surface of the film.
\par
The calculation of the RKKY interaction requires the real-space retarded Green's function. Starting from the Hamiltonian $H(\mathbf{k})$ given in Eq. (\ref{eq5}), we express the Green's function $G(\pm \mathbf{R}, \epsilon)$ in Lehmann's representation as
\begin{equation}\label{eq12}
G\left( \pm \mathbf{R},\epsilon \right) =\frac{1}{\left(2\pi\right)^2}\int e^{\pm i\mathbf{k\cdot R}}
\frac{1}{\epsilon+i0^+-H(\mathbf{k})}d^{2}\mathbf{k} .
\end{equation}%
Because $H(\mathbf{k})$ acts on a space that combines the degrees of freedom from both the top and bottom surfaces, the Green's function takes the following block-matrix form:
\begin{equation}\label{eq13}
G( \pm \mathbf{R},\epsilon )=\left(
\begin{array}{cc}
G_{tt}( \pm \mathbf{R},\epsilon) & G_{tb}( \pm \mathbf{R},\epsilon ) \\
G_{bt}( \pm \mathbf{R},\epsilon) & G_{bb}( \pm  \mathbf{R},\epsilon)%
\end{array}%
\right) ,
\end{equation}%
where the subscript $t$ ($b$) indicates that the impurity is located on the top (bottom) surface of the film.
\par
In this work, we consider two distinct impurity configurations: both impurities on the same surface, and one impurity on the top surface with the other on the bottom. We begin by examining the former scenario and, for convenience, assume that both impurities reside on the top surface. In this case, substituting $H(\mathbf{k})$ from Eq. (\ref{eq5}) into Eq. (\ref{eq12}) and performing algebraic manipulations yields $G_{tt}(\pm \mathbf{R},\epsilon)$ in the form
\begin{equation}\label{eq14}
G_{tt}\left( \pm \mathbf{R},\epsilon\right) =\left(
\begin{array}{cc}
f_{0}+f_{z} & \pm e^{-i\theta _{R}}f \\
\mp e^{i\theta _{R}}f & f_{0}-f_{z}%
\end{array}%
\right) .
\end{equation}
The matrix elements ($f_0$, $f_z$, $f$) of $G_{tt}(\pm \mathbf{R},\epsilon)$ depend on the SL count via $\lambda$ and are given explicitly by
\begin{eqnarray}\label{eq15}
\begin{split}
f_{0}\left( \lambda =-\right) & =-\underset{s^{\prime }=\pm }{\sum }\frac{ \epsilon(\gamma +s^{\prime }\eta mm_{\omega
})K_{0}\left( R/\sqrt{\frac{v^{2}}{\zeta _{- ,s^{\prime
	}}^{2}-\epsilon ^{2}}}\right)}{\gamma/\alpha } , \\
f_{z}\left( \lambda =-\right) & =-\underset{%
s^{\prime }=\pm }{\sum }\frac{\alpha \left[ (m+\eta m_{\omega })(\gamma +s^{\prime }\eta mm_{\omega
})+s^{\prime }\eta m_{\omega }\Delta ^{2}\right]}{\gamma / K_{0}\left( R/\sqrt{\frac{v^{2}}{\zeta _{- ,s^{\prime
}}^{2}-\epsilon ^{2}}}\right)} , \\
f\left( \lambda =-\right) & =-\frac{\alpha }{\gamma }\underset{s^{\prime
}=\pm }{\sum }\frac{\left( \gamma +s^{\prime }\eta mm_{\omega }\right) }{\sqrt{%
\zeta _{- ,s^{\prime }}^{2}-\epsilon ^{2}}}K_{1}\left( R/\sqrt{\frac{v^{2}}{\zeta _{- ,s^{\prime
}}^{2}-\epsilon ^{2}}}\right) ,\\
f_{0}\left( \lambda =+\right) & =-\alpha \epsilon \underset{s^{\prime
}=\pm }{\sum }K_{0}\left( R/\sqrt{\frac{v^{2}}{\zeta _{+ ,s^{\prime
}}^{2}-\epsilon ^{2}}}\right) , \\
f_{z}\left( \lambda =+\right) & =-\alpha \underset{s^{\prime }=\pm }{\sum
}\zeta _{+ ,s^{\prime }}K_{0}\left( R/\sqrt{\frac{v^{2}}{\zeta _{+ ,s^{\prime}}^{2}-\epsilon ^{2}}}\right) , \\
f\left( \lambda =+\right) & =-\alpha \underset{s^{\prime }=\pm }{\sum }\frac{%
1}{\sqrt{\frac{1}{\zeta _{+ ,s^{\prime }}^{2}-\epsilon ^{2}}}}%
K_{0}\left( R/\sqrt{\frac{v^{2}}{\zeta _{+ ,s^{\prime}}^{2}-\epsilon ^{2}}}\right) ,
\end{split}%
\end{eqnarray}%
where $\alpha =1 /4\pi v^{2}$,  $\gamma =m_{\omega }\sqrt{%
\Delta ^{2}+m^{2}}$, $\zeta _{+,s^{\prime }}=m+\eta m_{\omega }+s^{\prime }\Delta $, $\zeta _{-,s^{\prime }}=\left|  m_{\omega }+s^{\prime }\sqrt{\Delta ^{2}+m^{2}}\right| $ and $K_n(x)$ ($n=0,1$) denotes the $n$th-order modified Bessel function of the second kind.
\par
By substituting $G_{tt}(\pm \mathbf{R},\epsilon)$ from Eq. (\ref{eq14}) into Eq. (\ref{eq11}) and tracing over the spin degrees of freedom, the RKKY interaction $H_{R}^{tt}$ can be expressed in the following form:
\begin{equation}\label{eq16}
H_{R}^{tt}(\lambda)=\underset{i}{\sum }%
J_{ii}(\lambda)S_{1}^{i}S_{2}^{i}+J^z_f(\lambda)(S_{1}^{x}S_{2}^{y}+S_{1}^{y}S_{2}^{x})+J_{DM}(\lambda)%
\widetilde{\mathbf{e}}_{R}\cdot (\mathbf{S}_{1}\times \mathbf{S}_{2}),
\end{equation}
with
\begin{eqnarray}\label{eq17}
\begin{split}
J_{xx}\left( \lambda \right) & =-\frac{2J_{c}^{2}}{\pi }\mathrm{Im}%
\int_{-\infty }^{\epsilon _{F}}[f_{0}^{2}-f_{z}^{2}-f^{2}\cos (2\theta
_{R})]d\epsilon , \\
J_{yy}\left( \lambda \right) & =-\frac{2J_{c}^{2}}{\pi }\mathrm{Im}%
\int_{-\infty }^{\epsilon _{F}}[f_{0}^{2}-f_{z}^{2}+f^{2}\cos (2\theta
_{R})]d\epsilon , \\
J_{zz}\left( \lambda \right) & =-\frac{2J_{c}^{2}}{\pi }\mathrm{Im}%
\int_{-\infty }^{\epsilon _{F}}[f_{0}^{2}+f_{z}^{2}-f^{2}]d\epsilon , \\
J^z_{f}\left( \lambda \right) & =-\frac{2J_{c}^{2}}{\pi }\mathrm{Im}%
\int_{-\infty }^{\epsilon _{F}}[-f^{2}\sin (2\theta _{R})]d\epsilon , \\
J_{\text{DM}}\left( \lambda \right) & =-\frac{4J_{c}^{2}}{\pi }\mathrm{Im}%
\int_{-\infty }^{\epsilon _{F}}\left(f_{0}f\right) d\epsilon
\end{split}%
\end{eqnarray}%
where $\widetilde{\mathbf{e}}_{R}=(-\sin \theta _{R},\cos \theta _{R},0)$. In Eq. (\ref{eq16}), $J_{ii}$ couples collinear spins, $J^z_{f}$ represents the spin-frustrated term, and $J_{\text{DM}}$ corresponds to the Dzyaloshinskii-Moriya (DM) interaction.
\par
For the configuration where the two impurities are placed on the top and bottom surfaces respectively, the corresponding Green's function $G_{bt}\left( \mathbf{R},\epsilon\right)$ takes the form
\begin{equation}\label{eq18}
G_{bt}\left( \mathbf{R},\epsilon\right) =\left(
\begin{array}{cc}
g_{0}+g_{z} & e^{-i\theta _{R}}g\\
\lambda e^{i\theta _{R}}g & g_{0}-g_{z}%
\end{array}%
\right),
\end{equation}%
where $g_0$, $g_z$, and $g$ are functions of $\lambda$, given explicitly by
\begin{eqnarray}\label{eq19}
\begin{split}
g_{0}\left( \lambda =-\right) & =-\frac{\alpha \Delta v^{2}/\hslash \omega}{ \sqrt{%
\Delta ^{2}+m^{2}} } \underset{%
s^{\prime }=\pm }{\sum }(k_0^2 +s^{\prime }k_{a}^{2})K_{0}\left( R/\sqrt{\frac{v^{2}}{\zeta _{- ,s^{\prime}}^{2}-\epsilon ^{2}}}\right) , \\
g_{z}\left( \lambda =-\right) & =-\frac{\alpha }{\gamma }\eta \Delta m_{\omega
}\epsilon \underset{s^{\prime }=\pm }{\sum }s^{\prime }K_{0}\left( R/\sqrt{\frac{v^{2}}{\zeta _{- ,s^{\prime}}^{2}-\epsilon ^{2}}}\right) , \\
g\left( \lambda =-\right) & =\frac{\alpha }{\gamma }\eta \Delta m_{\omega }%
\underset{s^{\prime }=\pm }{\sum }\frac{s^{\prime }K_{1}\left( R/\sqrt{\frac{v^{2}}{\zeta _{- ,s^{\prime}}^{2}-\epsilon ^{2}}}\right)}{\sqrt{\frac{1}{%
\zeta _{- ,s^{\prime }}^{2}-\epsilon ^{2}}}} , \\
g_{0}\left( \lambda =+\right) & =-\alpha \underset{s^{\prime }=\pm }{\sum
}s^{\prime }\zeta _{+ ,s^{\prime }}K_{0}\left( R/\sqrt{\frac{v^{2}}{\zeta _{+ ,s^{\prime}}^{2}-\epsilon ^{2}}}\right) , \\
g_{z}\left( \lambda =+\right) & =-\alpha \epsilon \underset{s^{\prime
}=\pm }{\sum }s^{\prime }K_{0}\left( R/\sqrt{\frac{v^{2}}{\zeta _{+ ,s^{\prime}}^{2}-\epsilon ^{2}}}\right) , \\
g\left( \lambda =+\right) & =-\alpha \underset{s^{\prime }=\pm }{\sum }\frac{%
s^{\prime }K_{1}\left( R/\sqrt{\frac{v^{2}}{\zeta _{+ ,s^{\prime}}^{2}-\epsilon ^{2}}}\right)}{\sqrt{\frac{1}{\zeta _{+ ,s^{\prime }}^{2}-\epsilon
^{2}}}} .
\end{split}%
\end{eqnarray}%
Following the same procedure as for $H_{R}^{tt}$, we obtain the RKKY interaction $H_{R}^{tb}$ as
\begin{equation}\label{eq20}
H_{R}^{tb}(\lambda)=\underset{i}{\sum }\mathcal{J}_{ii}(\lambda)S_{1}^{i}S_{2}^{i}+%
\sum_{i=x,y,z}\mathcal{J}_{f}^{i}(\lambda)(S_{1}^{j}S_{2}^{k}+S_{1}^{k}S_{2}^{j}),
\end{equation}%
where $(j,k)$ form an even permutation of the Levi-Civita symbol for a fixed $i$. Unlike $H_{R}^{tt}$ in Eq. (\ref{eq16}), $H_{R}^{tb}$ lacks the DM interaction but contains two additional spin-frustrated terms, $\mathcal{J}_f^x$ and $\mathcal{J}_f^y$. Their explicit expressions, together with the other components, are
\begin{equation}\label{eq21}
\begin{split}
\mathcal{J}_{xx}\left( \lambda \right) & =-\frac{2J_{c}^{2}}{\pi }\mathrm{Im}%
\int_{-\infty }^{\epsilon _{F}}[g_{0}^{2}-g_{z}^{2}+g^{2}\cos (2\theta
_{R})]d\epsilon , \\
\mathcal{J}_{yy}\left( \lambda \right) & =-\frac{2J_{c}^{2}}{\pi }\mathrm{Im}%
\int_{-\infty }^{\epsilon _{F}}[g_{0}^{2}-g_{z}^{2}-g^{2}\cos (2\theta
_{R})]d\epsilon , \\
\mathcal{J}_{zz}\left( \lambda \right) & =-\frac{2J_{c}^{2}}{\pi }\mathrm{Im}%
\int_{-\infty }^{\epsilon _{F}}[g_{0}^{2}+g_{z}^{2}-g^{2}]d\epsilon , \\
\mathcal{J}_{f}^{x}\left( \lambda  \right) & =-\frac{4J_{c}^{2}}{\pi }\mathrm{Im}%
\int_{-\infty }^{\epsilon _{F}}\left[\theta(\lambda)g_z +\theta(-\lambda)g_0  \right]g  \sin \theta _{R}d\epsilon , \\
\mathcal{J}_{f}^{y}\left( \lambda  \right) & =-\frac{4J_{c}^{2}}{\pi }\mathrm{Im}%
\int_{-\infty }^{\epsilon _{F}}\left[\theta(\lambda) g_z +\theta(-\lambda)g_0 \right]g  \cos \theta _{R}d\epsilon , \\
\mathcal{J}_{f}^{z}\left( \lambda \right) & =-\frac{2J_{c}^{2}}{\pi }\mathrm{Im}%
\int_{-\infty }^{\epsilon _{F}}\left[ g^{2}\sin (2\theta _{R})\right] d\epsilon ,
\end{split}%
\end{equation}%
where $\theta(\lambda)$ denotes the Heaviside step function.

\section{Results and discussions}
\label{results}
This section is structured into three parts. In the first two parts, we focus on the RKKY interactions in the absence of light. Specifically, the first part presents the magnetic signatures that distinguish MnBi$_2$Te$_4$ films ($m \neq 0$) from nonmagnetic topological insulator films ($m = 0$). The second part identifies the magnetic signals that differentiate AI phase (even-SL) and QAH phase (odd-SL). In the final part, we analyze RKKY interactions under off-resonant CPL. Here, by monitoring the evolution of the RKKY interaction with light, we extract characteristic magnetic signatures of the topological phase transitions. These film-type-dependent signatures are expected to provide further discrimination between AI phase (even-SL) and QAH phase (odd-SL).
\subsection{Distinguishing MnBi$_2$Te$_4$ ($m \neq 0$) and nonmagnetic topological insulator ($m = 0$)}
\begin{figure}[tbh]
\centering \includegraphics[width=0.35\textwidth]{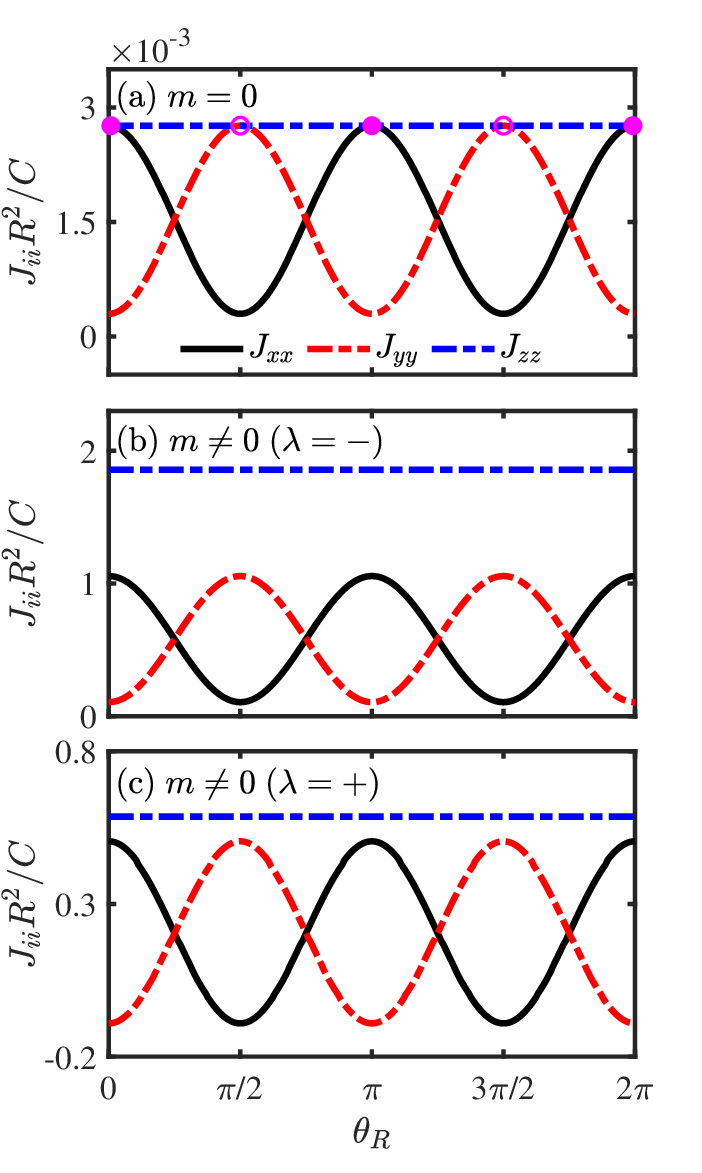}
\caption{Dependence of the collinear RKKY components $J_{ii}$ on the azimuthal angle $\theta_R$ of the impurities, with impurities placed on the same surface. Results are presented for (a) $m=0$, (b) $m=0.025$ eV with $\lambda=-$ (even-SL films), and (c) $m=0.025$ eV with $\lambda=+$ (odd-SL films). Other parameters are $\epsilon_F = 0.04$ eV, $\Delta = 0.02$ eV, $v = 2.95$ eV$\cdot$\AA, $a = \Delta / v$, and $Ra =10$.}
\label{fig6}
\end{figure}
We begin by examining the collinear RKKY components $J_{ii}$, as they are particularly sensitive to the presence of finite $m$. For impurities deposited on the same surface with a fixed separation $R$, the dependence of $J_{ii}$ on the azimuthal angle $\theta_R =\arctan(R_y / R_x)$ is computed using Eqs.~\eqref{eq15} and~\eqref{eq17}, as summarized in Fig.~\ref{fig6}. In the case of $m = 0$, the results in Fig.~\ref{fig6}(a) align with previous work \cite{55}. In particular, when impurities are positioned along the axial directions $\theta_R = n\pi/2$ ($n=0,1,2,3$), the components $J_{ii}$ satisfy $J_{zz} = J_{xx} \neq J_{yy}$ (solid circles in Fig.~\ref{fig6}(a)) or $J_{zz} = J_{yy} \neq J_{xx}$ (open circles), corresponding to $XYX$-type and $XYY$-type RKKY spin models, respectively. It is noteworthy that both models exhibit a moderate anisotropy, governed primarily by the $f^2$ term in Eq. (\ref{eq17}).

In contrast, under a finite $m$, both even-SL ($\lambda=-$ in Fig. \ref{fig6}(b)) and odd-SL ($\lambda=+$ in Fig. \ref{fig6}(c)) display the same qualitative behavior: the original $XYX$-type (at $\theta_R = 0,\pi$) or $XYY$-type (at $\theta_R = \pi/2,3\pi/2$) model transforms into an $XYZ$-type model, characterized by $J_{xx} \neq J_{zz} \neq J_{yy}$. This $XYZ$ spin model exhibits the strongest anisotropy---a stark contrast to the moderate anisotropy found at $m = 0$. The emergence of such extreme anisotropy is entirely due to the finite $m$ (i.e., intrinsic magnetism), which introduces an additional correction to $J_{ii}$ via the $f_z^2$ term in Eq. (\ref{eq17}). Notably, the spin model originating purely from $f_z^2$ is of $XXZ$-type, distinct from the $f^2$-induced $XYX$- or $XYY$-type models. Thus, the presence of finite $m$ leads to a hybrid of $XXZ$ and $XYX$ (or $XYY$), resulting in the common $XYZ$-type spin model seen in both Fig. \ref{fig6}(b) and (c).

Taken together, these findings confirm that the RKKY spin model establishes a clear diagnostic criterion, i.e., the marked anisotropy contrast of this spin model between the magnetic ($m \neq 0$) and non-magnetic ($m = 0$) cases, for determining the introduction of intrinsic magnetism. Thus, it robustly distinguishes MnBi$_2$Te$_4$ ($m\neq0$) from nonmagnetic topological insulators ($m=0$). 

\subsection{Differentiating AI phase (even-SL) and QAH phase (odd-SL) in the dark}
The preceding discussion reveals a key limitation: within the $m\neq0$ framework, the spin model exhibits universality across AI phase (even-SL) and QAH phase (odd-SL), and fails to effectively differentiate between them, as evidenced by the identical $XYZ$-type spin model in Fig. \ref{fig6}(b,c). This limitation originates from the inability of the spin model to capture the detailed band properties of MnBi$_2$Te$_4$ films shown in Figs. \ref{fig1} and \ref{fig2}. To distinguish between AI phase (even-SL) and QAH phase (odd-SL), one must turn to magnetic signals that are sensitive to these band properties.
\par
\subsubsection{Characteristic Kinks: $\epsilon_F$-Dependent RKKY Interaction}
\begin{figure}[tbh]
\centering \includegraphics[width=0.35\textwidth]{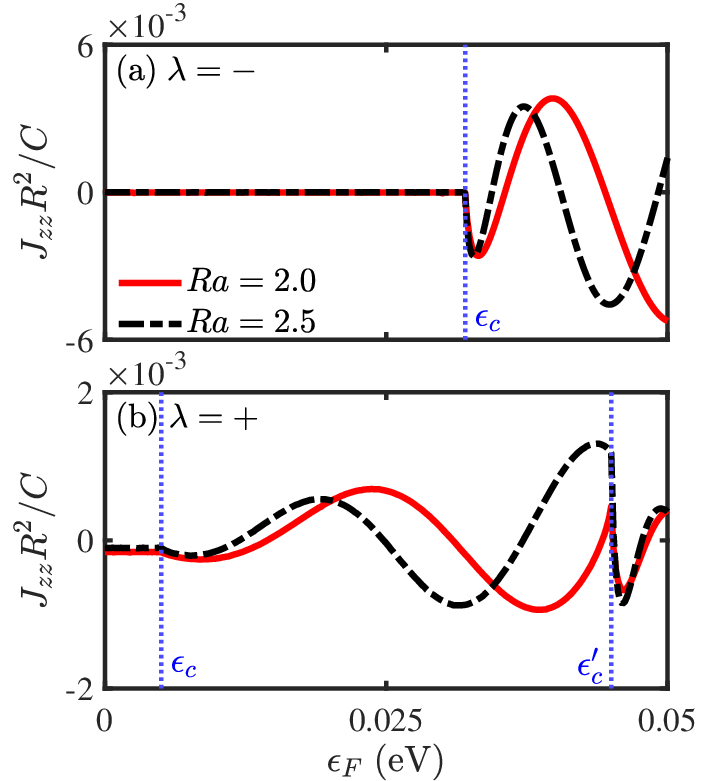}
\caption{The RKKY component $J_{zz}$ as a function of the Fermi energy $\epsilon_F$, with impurities placed on the same surface. Results are presented for (a) even-SL ($\lambda = -$) and (b) odd-SL ($\lambda = +$) films with different impurity distances ($Ra=2.0,2.5$). Other parameters are $\Delta=0.02$ eV, $v=2.95$ eV$\cdot $\AA , $a=\Delta /v$ and $\protect\theta _{R}=\protect\pi /4$.}
\label{fig7}
\end{figure}

An effective approach is to investigate the evolution of $J_{zz}$ with the Fermi energy $\epsilon_F$, as shown in Fig.~\ref{fig7}. In both cases ($\lambda=\pm$), a primary kink is observed at $\epsilon_c = \xi_g(\lambda)/2$, where $\xi_g(\lambda)$ is the band gap for the even- ($\lambda=-$) and odd-SL ($\lambda=+$) films in Fig.~\ref{fig1}. This kink arises because the Fermi energy $\epsilon_F$ crosses the band edge: when $\epsilon_F < \epsilon_c$, it lies within the band gap, suppressing $J_{zz}$; when $\epsilon_F > \epsilon_c$, electrons from the conduction band activate the interaction, causing $J_{zz}$ to rise abruptly. Thus, this kink serves as a universal signature of the band gap $\xi_g(\lambda)$.
\par
The critical distinction, however, emerges for the $\lambda=+$ case, where a second, distinctive kink is observed at a higher energy $\epsilon^\prime_c =  \xi^\prime_g(\lambda=+)/2$, as seen in Fig.~\ref{fig7}(b). This kink is determined by the gap $ \xi^\prime_g(\lambda=+)$ between the bands $ \xi_{+,+}$ and $ \xi_{-,+}$ in Fig.~\ref{fig1}(b). Its physical origin lies in the unique band splitting of the $\lambda=+$ case. Mechanistically, when $\epsilon_F < \epsilon^\prime_c$, the magnetic interaction originates only from the $ \xi_{+,-}$ band. Once $\epsilon_F$ surpasses $\epsilon^\prime_c$, electrons from the split band $ \xi_{+,+}$ begin to contribute, causing a sudden change in $J_{zz}$ that manifests as the kink. This second kink, directly linked to the SL-specific band splitting, therefore provides a definitive magnetic signature for distinguishing QAH phase (odd-SL) from AI phase (even-SL). Furthermore, this signal is robust, as the kink positions are independent of the impurity distance $R$ (as shown in Fig.~\ref{fig7}) and are also observable in other RKKY components (not shown here).
\par
\begin{figure}[tbh]
\centering \includegraphics[width=0.35\textwidth]{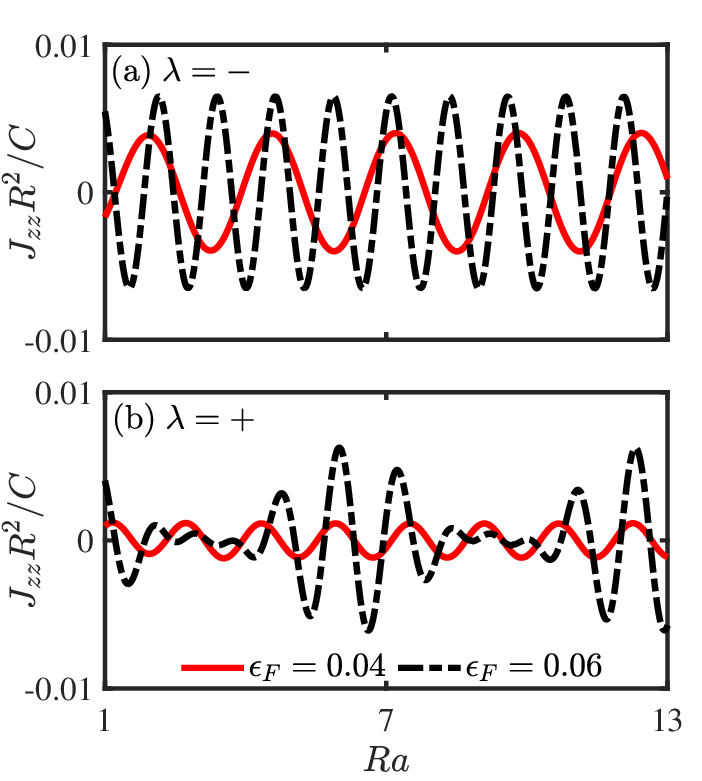}
\caption{The $R$-dependent RKKY component $J_{zz}$ for (a) even-SL ($\lambda = -$) and (b) odd-SL ($\lambda = +$) films, with impurities placed on the same surface. Different Fermi energies ($\epsilon_F = 0.04, 0.06$ \text{ eV}) are considered, and other parameters are identical to those in Fig. \ref{fig7}.}
\label{fig8}
\end{figure}

\subsubsection{Distinct Oscillation Patterns}
Alternatively, the AI phase (even-SL) and QAH phase (odd-SL) can also be distinguished by investigating the oscillatory behavior of $J_{zz}$ as a function of the impurity distance $R$. As shown in Fig. \ref{fig8}(a), for the $\lambda = -$ case, $J_{zz}$ always exhibits a single-period oscillation, regardless of the Fermi energy $\epsilon_F$. This behavior is dictated by the Fermi surface of the even-SL films. As depicted in Fig. \ref{fig2}(a, b), their Fermi surface consistently consists of a single contour. This means that changing $\epsilon_F$ only alters the size of this contour without changing its topological nature. Consequently, a single Fermi contour has a single radius and thus corresponds to a single Fermi wave number $k_F$ (Fig. \ref{fig2}(b)), which in turn directly dictates a single oscillation period ( $T = \pi / k_F$) for the interaction \cite{56,57,58}.
\par
In contrast, for the $\lambda = +$ (odd-SL) case, $J_{zz}$ maintains a single-period oscillation at lower Fermi energies but develops a double-period oscillation at larger values, as shown in Fig. \ref{fig8}(b). The emergence of this double-period oscillation stems from the peculiar Fermi surface of odd-SL films at large Fermi energies, which comprises two separated concentric contours (Fig. \ref{fig2}(d)). Their distinct radii correspond to two different Fermi wave numbers, $k_{F_-}$ and $k_{F_+}$ in Fig. \ref{fig2}(d), which naturally provide two distinct oscillation periods ($\pi / k_{F_-}$ and $\pi / k_{F_+}$) for the magnetic interaction. This transition from single- to double-period oscillation precisely reflects a topological deformation of the Fermi surface---known as a Lifshitz transition---in the odd-SL films. Hence, by tracing the evolution of the oscillation pattern with different $\epsilon_F$, one can effectively distinguish between the AI phase (even-SL) and QAH phase (odd-SL).
\par
\subsubsection{Presence/Absence of Spin-Frustrated Terms}
All magnetic signals discussed previously were obtained with impurities placed on the same surface. In contrast, placing impurities on different surfaces yields distinct magnetic signals, which can also distinguish between AI phase (even-SL) and QAH phase (odd-SL). Due to the vanishing of $g_z(\lambda=-)$ and $g(\lambda=-)$ in Eq. (\ref{eq19}) in the absence of CPL, the RKKY interaction $H_{R}^{tb}$ for the even-SL ($\lambda=-$) case, given in Eq. (\ref{eq20}), simplifies to
\begin{equation}\label{eq22}
H_{R}^{tb}(\lambda=-)=\mathcal{J}_H \mathbf{S}_{1}\cdot \mathbf{S}_{2},
\end{equation}
where $\mathcal{J}_H$ originates exclusively from $g_0(\lambda=-)$. The above equation indicates that the interaction here is purely collinear, consisting solely of a Heisenberg term.

For the $\lambda=+$ case, however, the RKKY interaction retains the general form of Eq. (\ref{eq20}). It therefore includes not only collinear RKKY components but also three additional types of spin-frustrated terms (non-collinear components) $\mathcal{J}^{x,y,z}_f$, which are always absent in the $\lambda=-$ case [Eq. (\ref{eq22})]. The emergence of these additional terms is directly attributable to the band splitting in odd-SL ($\lambda=+$) films. This splitting forces $g(\lambda=+)$ in Eq. (\ref{eq19}) to be non-zero, which, according to Eq. (\ref{eq21}), is the necessary condition for generating the spin-frustrated terms. The non-zero nature of $g(\lambda=+)$ can be understood from its explicit form: $g(\lambda=+) \propto \sum_{s^\prime} s^\prime K_0(x_{s^\prime})$, where $s^\prime = \pm$ labels the two split bands $\xi_{+, s^\prime}$. This expression represents the difference between the contributions from these bands. Since band splitting implies $x_{+} \neq x_{-}$, it follows that $K_0(x_+) \neq K_0(x_-)$, thereby guaranteeing a non-zero result for the summation. Consequently, for impurities on different surfaces, the two types of phases (AI and QAH phases in even- and odd-SL films) can be unambiguously distinguished simply by observing the presence or absence of spin-frustrated terms.
\par
Collectively, unlike the spin model constructed within the $m\neq0$ framework in the subsection III-A---which fails to reflect the specific impact of intrinsic magnetism on the surface-state bands---the magnetic signatures extracted here can clearly delineate the precise modifications that intrinsic magnetism imposes on the surface-state bands of MnBi$_2$Te$_4$ films. These include corrections to the surface-state energy gap, the presence or absence of band splitting, and deformations of the Fermi surface. Therefore, the distinct RKKY signatures revealed here---such as the secondary kink, the transition in oscillation, and the emergence of spin-frustrated terms---can serve as a magnetic alternative for distinguishing between AI phase (even-SL) and QAH phase (odd-SL). This approach provides a transport-independent means to resolve the experimental ambiguities encountered in prior transport-based studies \cite{20,24}, where electrical measurements alone cannot reliably differentiate the AI phase (even-SL) from the QAH phase (odd-SL).

In addition, we believe the above approach based on RKKY interactions applies not only to MnBi$_2$Te$_4$ films but also to other material systems hosting the AI and QAH phases. For instance, Di Xiao \textit{et al.} realized the AI phase in a QAH sandwich heterostructure \cite{Xiao2018}. As noted in that work, a key condition for realizing the AI phase is that all surfaces are gapped with the chemical potential lying inside the gaps. This condition makes our approach applicable to their system as well. Thus, an approach similar to the one used in Fig.~7 (analyzing the RKKY interaction as a function of Fermi energy $\epsilon_F$) can also be used to distinguish the AI and QAHE phases in that system. Specifically, in the QAH phase of Ref.~\cite{Xiao2018}, the side-surface states remain gapless (supporting chiral edge modes), so no kink appears in the $\epsilon_F$-dependent RKKY amplitude. In contrast, in the AI phase, where the side-surface states are expected to be gapped under ideal conditions (via quantum confinement), a distinct kink emerges when $\epsilon_F$ sweeps across the gap boundary. This distinguishing signal stems from the essential difference in the gap states of the surfaces between the AI and QAH phases, which is universal across material systems and makes our approach broadly applicable.

\subsection{Differentiating AI phase (even-SL) and QAH phase (odd-SL) via phase-transition signatures under illumination}
\begin{figure}[tbh]
\centering \includegraphics[width=0.49\textwidth]{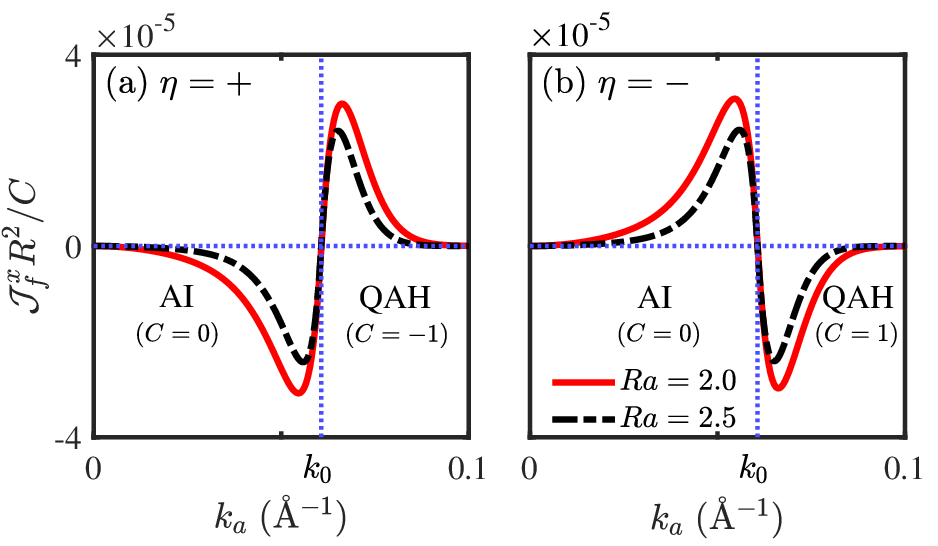}
\caption{The frustrated term $\mathcal{J}^{x}_{f}$ as a function of $k_{a}$ for
even-SL MnBi$_{2}$Te$_{4}\ $($\protect\lambda =-1$) under the (a) right-handed ($\eta=+$) and (b) left-handed ($\eta=-$) CPL, respectively. Impurities are placed on different surfaces with different distances ($Ra=2.0$, $2.5$). Here we set $\protect\epsilon _{F}=0$, $\Delta =0.02$ eV,
$v=2.95$ eV$\cdot $\AA , $a=\Delta /v$, $\protect\theta _{R}=%
\protect\pi /4$ and $m=0.025$ eV.}
\label{fig9}
\end{figure}
Beyond the magnetic signatures in unperturbed systems discussed above, we also analyze those induced by topological phase transitions under illumination, which provide additional information for distinguishing between AI phase (even-SL) and QAH phase (odd-SL). To this end, we examine the RKKY interaction under CPL, focusing on how its distinct components respond differently in even- and odd-SL films. The Fermi energy is set to zero to best capture the band gap evolution, and impurities are placed on different surfaces for maximum sensitivity to the transition.
\par
For even-SL ($\lambda=-$) films, the spin-frustrated term $\mathcal{J}^x_{f}$ (or $\mathcal{J}^y_{f}$) serves as a sharp diagnostic tool. As shown in Fig. \ref{fig9}, $\mathcal{J}^x_{f}$ undergoes a clear sign reversal precisely at the topological transition point $k_a=k_0$, with the direction of reversal (negative-to-positive or vice versa) dictated by the light's circular polarization chirality $\eta$ (Fig. \ref{fig9}). This $\eta$-dependent sign reversal directly maps onto the chirality-controlled transition between the AI ($C=0$) and QAH ($C=\pm1$) states depicted in Fig. \ref{fig5}(a). The origin of this unique signature lies in the analytical form $\mathcal{J}^x_{f} \propto \eta(k_a^2 - k_0^2)$ for $\lambda=-$ (Eqs. (\ref{eq19}) and (\ref{eq21})), which inherently couples the polarization chirality $\eta$ to the transition point $k_a=k_0$.
\begin{figure}[tbh]
\centering \includegraphics[width=0.49\textwidth]{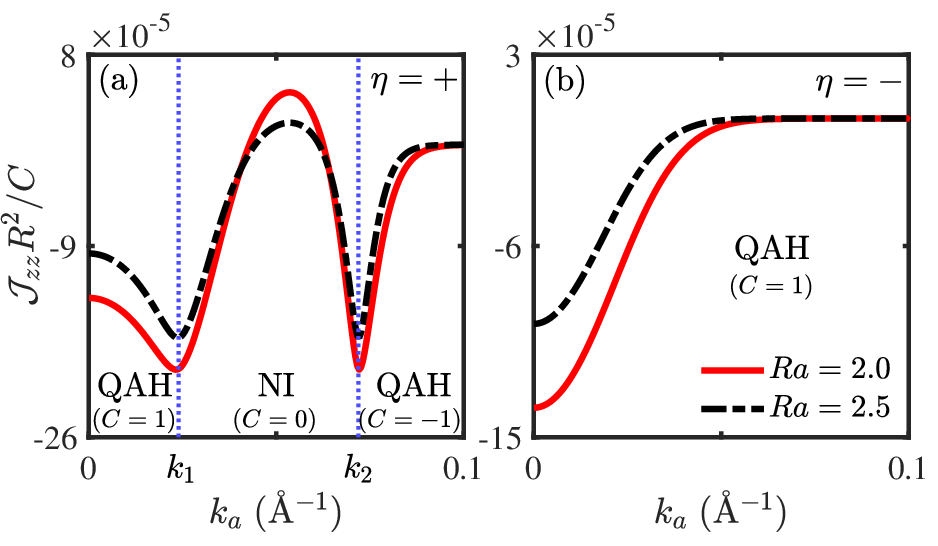}
\caption{The RKKY component $\mathcal{J}_{zz}$ as a function of $k_{a}$ for
odd-SL MnBi$_{2}$Te$_{4}\ $($\protect\lambda =+$) under the (a) right-handed ($\eta=+$) and (b) left-handed ($\eta=-$) CPL, respectively. Impurities are placed on different surfaces with different distances ($Ra=2.0$, $2.5$). Other parameters are identical to those in Fig. \ref{fig9}.}
\label{fig10}
\end{figure}
\par
In stark contrast, odd-SL ($\lambda=+$) films exhibit a completely different magnetic signature under CPL. Here, the collinear RKKY components $\mathcal{J}_{ii}$ shown in Fig. \ref{fig10}, rather than the spin-frustrated terms $\mathcal{J}^{x,y}_{f}$, serve as the key probes. Under right-handed ($\eta=+$) CPL, $\mathcal{J}_{zz}$ displays a characteristic double-dip structure as $k_a$ increases, with the two dips pinpointing the phase boundaries at $k_1$ and $k_2$ (Fig. \ref{fig10}(a)), corresponding directly to the two topological transitions shown in the phase diagram of Fig. \ref{fig5}(b). These dips arise because the RKKY interaction peaks when a band edge touches the Fermi level at the gap-closing points---a direct consequence of the enhanced scattering probability for electrons near the band edge \cite{59_1,59_2,59_3,59_4,59_5}. Under left-handed ($\eta=-$) CPL, however, $\mathcal{J}_{zz}$ monotonically decays without any dip structure (Fig. \ref{fig10}(b)), indicating the absence of a topological transition---a behavior again distinct from that of even-SL films and consistent with the constant topology in Fig. \ref{fig5}(b).

Taken together, our results reveal a clear diagnostic dichotomy under illumination: even-SL films exhibit chirality-dependent sign reversals in the spin-frustrated components ($\mathcal{J}^{x,y}_f$), while odd-SL films are characterized by chirality-selective double-dip structures in the collinear RKKY components. Given that the AI and QAH phases are the respective initial phases of even-SL and odd-SL films, this fundamental difference demonstrates that these phase-transition magnetic signatures provide a powerful probe to distinguish between AI phase (even-SL) and QAH phase (odd-SL), thereby extending their fingerprinting beyond unperturbed systems.

In addition, these signatures differ fundamentally from those reported in previous RKKY studies of light-driven transitions (such as Refs.~\cite{44_2,58_2}). In those works, observed sign reversals occur in the DM interaction, and only single-dip structures appear in other RKKY components---both features being independent of the light's chirality. This distinction arises from the different nature of the topological phase transitions involved. In our case, the transitions are chirality- and film-type-dependent, a direct consequence of being governed by the intrinsic magnetism of MnBi$_2$Te$_4$ films. This clearly differentiates our physical scenario from those considered in Refs.~\cite{44_2,58_2}, which are based on non-magnetic systems.

\section{Summary}
\label{summary}
We systematically investigated the RKKY interaction in MnBi$_2$Te$_4$ films. In the absence of external fields, several key magnetic signatures were identified. First, with axially arranged impurities, the RKKY spin model exhibits significantly stronger anisotropy in MnBi$_2$Te$_4$ than in nonmagnetic topological insulators, providing a clear distinguishing feature. Furthermore, we established three diagnostic signatures to differentiate between AI phase (even-SL) and QAH phase (odd-SL), which depend on the specific impurity configurations. The first two signatures, obtained with impurities on the same surface, are: (i) characteristic kinks in the Fermi-energy dependence of the RKKY amplitude (a single kink for even-SL vs. two for odd-SL); and (ii) distinct real-space oscillation patterns (a persistent single-period oscillation for even-SL vs. a transition to double-period oscillation for odd-SL). A third signature, observed with impurities on different surfaces, is the presence (odd-SL) or absence (even-SL) of spin-frustrated terms. All these signatures originate from the fundamental difference in band structures between even- and odd-SL films, a difference that is ultimately induced by the intrinsic magnetism. Finally, under off-resonant CPL, we extracted distinct magnetic responses from the spin-frustrated and collinear components, which provide film-type-dependent signatures for topological phase transitions and thereby offer additional information to distinguish between AI phase (even-SL) and QAH phase (odd-SL).
\par
Our work shows that measuring the RKKY interaction provides an effective alternative for characterizing band properties and phase transitions in MnBi$_{2}$Te$_{4}$ thin films, thereby offering a magnetic perspective to understand the influence of intrinsic magnetism on the surface-state band structure. The proposed scheme is feasible with existing techniques, such as spin-polarized scanning tunneling spectroscopy \cite{59,60}, capable of detecting magnetization curves of individual atoms \cite{60}, or electron spin resonance techniques combined with optical detection methods \cite{61}.

\acknowledgements
This work was supported by the National Natural Science Foundation of China (Grant Nos. 12104167, 12174121, 11904107, 11774100), by the Guangdong Basic and Applied Basic Research Foundation under Grant No. 2023B1515020050, by GDUPS (2017) and by Key Program for Guangdong NSF of China (Grant No. 2017B030311003).

\section*{Data Availability} 
The data that support the findings of this article are not publicly available. The data are available from the authors upon reasonable request.
%\bibliographystyle{apsrev4-2}
%\bibliography{reference}

%apsrev4-2.bst 2019-01-14 (MD) hand-edited version of apsrev4-1.bst
%Control: key (0)
%Control: author (8) initials jnrlst
%Control: editor formatted (1) identically to author
%Control: production of article title (0) allowed
%Control: page (0) single
%Control: year (1) truncated
%Control: production of eprint (0) enabled
%
\end{document}